\documentstyle[spie]{article} 
\input{psfig}   

\title{Coupling of intrinsic Josephson oscillations in layered superconductors 
by charge fluctuations} 

\author{Ch. Preis\supit{a}, Ch. Helm\supit{a},
J. Keller\supit{a}, A. Sergeev\supit{a} and R. Kleiner\supit{b}  
\skiplinehalf 
\supit{a}Institute of Theoretical Physics, University of Regensburg,  
D-93040 Regensburg, Germany 
\skiplinehalf 
\supit{b}Physical Institute III, University of Erlangen-N\"urnberg, 
D-91058 Erlangen, Germany 
}

\authorinfo{E-mail: christian.preis@physik.uni-regensburg.de}
\begin{document} 

\maketitle 

\begin{abstract}

The coupling of Josephson oscillations in layered superconductors 
is studied with help of a tunneling Hamiltonian
formalism. The general form of the current density across the barriers 
between the superconducting layers is derived. The induced charge
fluctuations on the superconducting layers lead to a coupling of the
Josephson oscillations in different junctions. A simplified set of equations
is then used to study the non-linear dynamics of the system. In particular the
influence of the coupling on the current-voltage characteristics is
investigated and upper limits for the coupling strength are estimated from a
comparison with experiments on cuprate superconductors. 

\end{abstract} 


%%%%%%%%%%%%%%%%%%%%%%%%%%%%%%%%%%%%%%%%%%%%%%%%%%%%%%%%%%%%%

\section{INTRODUCTION}

\label{sect:intro}  

The superconducting properties of the highly anisotropic 
cuprate--superconductors Tl$_2$Ba$_2$Ca$_2$Cu$_3$O$_{10+\delta}$ (TBCCO) and
Bi$_2$Sr$_2$CaCu$_2$O$_{8+\delta}$ (BSCCO) 
 are well described  by 
a stack of Josephson junctions coupling the superconducting  CuO$_2$ 
layers  in  $c$-direction.    In  particular  the  multiple  branch 
structure  observed  in  the  current-voltage  characteristics  by 
several groups 
\cite{Kleiner92,Kleiner94,Schlenga96,Yurgens96,Regy94,Tanabe96,Itoh97} 
can be explained by this model. 

Due to the low value of the critical  current in $c$-direction  the 
system has a small Josephson  plasma frequency  $\omega_p$ 
and a 
large value of the McCumber  parameter  $\beta_c$,  which causes a 
strong    hysteretic    behaviour    of    the    current-voltage 
characteristics.   The low value of the Josephson plasma frequency 
manifests  itself  in the transparency  of the stack with respect 
to THz radiation  in $c$-direction \cite{Tajima93}.   
The longitudinal and transversal plasma oscillations have also been observed 
directly \cite{Uchida97,Kadowaki97}. 

The theoretical investigation of longitudinal plasma oscillations has become 
popular since Koyama and Tachiki \cite{Koyama96} 
proposed a coupling of Josephson 
oscillations  in different  barriers due to charge fluctuations. 
In systems  with weakly coupled  superconducting  layers the 
charges  on different layers need  not  to be constant, which is  in contrast  
to  ordinary  superconductors  where  charge  neutrality  can  be 
assumed.  In the theory of Koyama and Tachiki \cite{Koyama96} it follows 
that the gauge-invariant scalar potential 
\begin{equation}
\mu_l = \Phi_l - (\hbar/2e) \dot \chi_l
\end{equation}
does not vanish. Here $\Phi_l$ is the electric scalar potential and $\chi_l$ 
is the phase  of the  superconducting  order  parameter $\Delta_l = 
\vert \Delta_l\vert \exp(i\chi_l)$
on layer $l$.  On the other hand the Josephson current density $j_c\sin\gamma_{l,l+1}$
between layers  $l$ and 
$l+1$ depends on the  gauge-invariant phase difference 
\begin{equation}
\gamma_{l,l+1}(t) = \chi_l(t) - \chi_{l+1}(t) - {2e\over \hbar} \int_l^{l+1} dz
\, A_z(z,t) 
\end{equation}
where $A$ is the vector potential (in our notation $e=\vert e\vert$). 
Its time derivative (the 
second Josephson relation) 
\begin{equation}
{\hbar \over 2e} \dot \gamma_{l,l+1}(t) = \int_l^{l+1} dz\, E_z + \mu_{l+1} - \mu_l
\end{equation}
then not only depends  on the voltage between  the 
layers, but also on the potential difference $\mu_{l+1} - \mu_l$. 
This  finally  leads  to a coupling  between  Josephson 
oscillations in different barriers. 
A non-vanishing 
generalized  scalar potential which is related to  
quasi-particle charge   imbalance  can also be obtained 
without   breaking   charge 
neutrality.   The importance of this effect  has been stressed 
by Artemenko and Kobelkov \cite{Artemenko97} and by Ryndyk 
\cite{Ryndyk97,Ryndyk98}. 
In the layered cuprate-superconductors
probably both effects are present. 

In this communication  we want to derive the coupling effect in   
a  microscopic   model  starting  from  a  tunneling  Hamiltonian.
We arrive at an expression  for the current between different layers 
which  is  formally  similar  to that  obtained  by Artemenko  for  a 
different  model. In leading order in the interlayer hopping $t_\perp$ our
results are also similar to the model of 
Koyama and Tachiki.  This simplified model will be used to 
study  the  non-linear  dynamics of the system  and  to discuss 
implications of the coupling on the current-voltage characteristics.

\section{General outline of the theory}
\label{sect:theory}

We start from a model where the current between superconducting layers
across the insulating barrier is described by a time-dependent 
tunneling Hamiltonian 
\begin{equation}
H_T = \sum_{l,k,k',\sigma} T_{k,k'} c^\dagger_{l+1,k',\sigma}
c^{\phantom\dagger}_{l,k,\sigma} e^{-{ie\over \hbar}\int_l^{l+1} 
dz\, A_z(t)} +  h.c.
\end{equation}
which depends on the vector potential $A_z(t)$ in the barrier. 
Here $T_{k,k'}$ is a tunneling matrix element describing (random) 
hopping between neighboring layers. In order to get a Josephson current 
also for d-wave superconductors we have to keep some angular dependence in
this matrix element. 

In addition to this the current is driven by the difference of scalar
potentials $\Phi_{l}(t)$ on the different layers. 
Thus the total time-dependent Hamiltonian is 
\begin{equation}
H=\sum_l (H_l - e\Phi_l(t) N_l) + H_T(t)
\end{equation}
where $H_l$ is the Hamiltonian of the 
electrons  in layer $l$ including superconducting 
interactions and pairing. 

Finally we have to take into account phase fluctuations 
of the superconducting order parameter induced by charge fluctuations. 
It can be shown that the results of the calculation for physical quantities 
depend only on the gauge-invariant combinations $\gamma_{l,l+1}$ and $\mu_l$ 
of the electromagnetic potentials with the phase of the order parameter.
Formally the same results are obtained if we replace in
the Hamiltonian the exponential by $\exp(i\gamma_{l,l+1}(t))$, the 
scalar potential $\Phi_l(t)$ by $\mu_l(t)$ and assume the order parameter 
to be real in the BCS-treatment of $H_l$. 

The non-linear Josephson effect results from the periodic  
dependence on the phases $\gamma(t)$. In the limit of large McCumber
parameter $\beta_c$ the phase can be written as 
\begin{equation}
\gamma(t) = \gamma_0 + \omega t + \delta\gamma (t)
\end{equation}
where $\gamma_0$ is the constant phase determined by the dc-current and  
$\omega$ is the Josephson frequency which is
related to the dc-voltage. The part $\delta \gamma(t)$, which 
oscillates with the same frequency $\omega$, is small for large $\beta_c$.

We calculate the current response $j_{l,l+1}$ between neighboring 
superconducting   layers  with  respect  to  both  the  tunneling 
Hamiltonian $H_T$ and the generalized scalar potential $\mu_l$ restricting 
ourselves to second order tunneling processes and linear response 
with respect to $\mu_l$. 

\begin{figure}
   \begin{center}
   \begin{tabular}{c}
   \psfig{figure=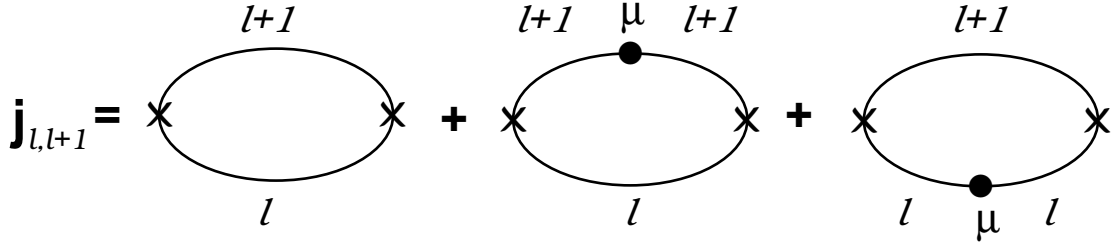} 
   \end{tabular}
   \end{center}
   \caption[example] 
   { \label{fig1}          
Graphs for the current density. Symbols:
left {\Large$\times$} =  current operator, right {\Large$\times$} =  $\rm H_T$,
{\Large$\bullet$} = density vertex.
Each cross corresponding to a hopping $T_{kk'}$ between
layers $l$ and $l+1$ is combined with a phase factor 
$\exp\left(\pm i\gamma_{l,l+1}(t)/2\right)$. 
} 
\end{figure}

\begin{figure}
   \begin{center}
   \begin{tabular}{c}
   \psfig{figure=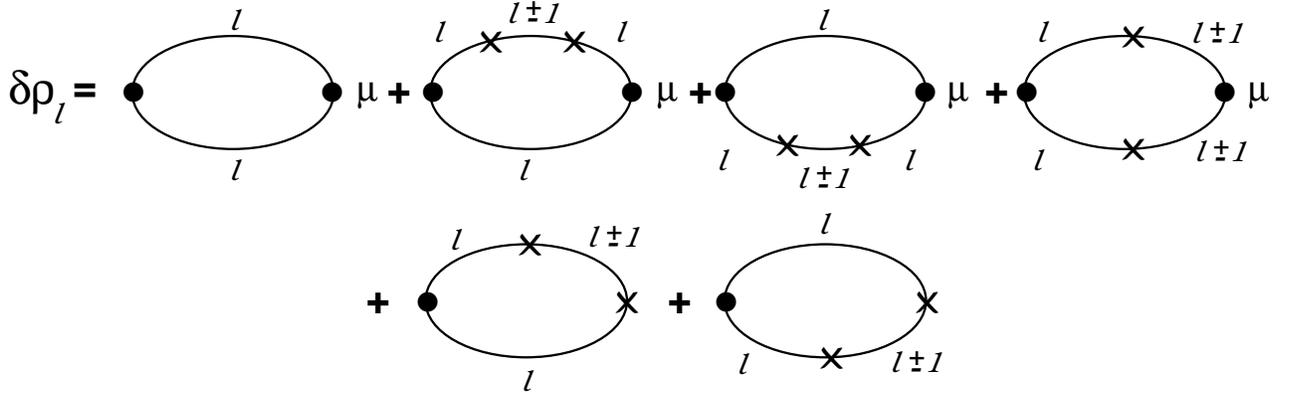,width=17cm} 
   \end{tabular}
   \end{center}
   \caption[example] 
   { \label{fig2}          
Graphs for the density response in layer $l$. 
Symbols:
{\Large$\times$} =  $\rm H_T$,
{\Large$\bullet$} = density vertex.
} 
\end{figure}

The result can be written in the following general form:
\begin{equation}
j_{l,l+1}(t) = j^{qp}_{l,l+1}(t) + j^J_{l,l+1}(t) 
\end{equation}

where
\begin{equation}
j^{qp}_{l,l+1}(t) = \int^t_{-\infty}  dt_1\, S_0(t-t_1) 
{\dot \gamma(t_1)\over 2} \cos{  \gamma(t)  -\gamma(t_1)\over  2}
+ \int_{-\infty}^t  dt_1 \int_{-\infty}^t  dt_2\, S_1(t,t_1,t_2)  
\cos {\gamma(t) - \gamma(t_1)\over  2}  (\mu_{l+1}(t_2)  - \mu_l(t_2))  
\end{equation} 
is the quasi-particle current density and 
\begin{eqnarray}
j^J_{l,l+1}(t)  & = & 
j_c \sin \gamma(t) + \int^t_{-\infty} 
dt_1\, J_0(t-t_1) {\dot \gamma(t_1)\over 2}  
\cos{\gamma(t)   +\gamma(t_1)\over   2}\nonumber\\
& + &     \int_{-\infty}^t    dt_1    \int_{-\infty}^t    dt_2  
\, J_1(t,t_1,t_2)    \cos    {\gamma(t)    +    \gamma(t_1)\over    2} 
(\mu_{l+1}(t_2) - \mu_l(t_2)) 
\end{eqnarray}
is the Josephson current carried by the condensate 
($\gamma(t):=\gamma_{l,l+1}(t)$). The functions 
$S_0(t-t_1)$ and $J_0(t-t_1)$ result from a folding of two normal and 
anomalous Green's functions in neighboring layers (see Fig. \ref{fig1}),
and $j_c = J_0(t,t)=const $ 
is the critical current density. 
In the terms $S_1(t,t_1,t_2)$ and 
$J_1(t,t_1,t_2)$  the additional linear dependence of the Green's 
functions on $\mu_l(t_2)$ is considered. 
A similar  (but more complicated  expression) is obtained  for the 
charge density response (see Fig. \ref{fig2}). 
Equations with a similar structure have been obtained by Artemenko and 
Kobelkov \cite{Artemenko97} for a different model. 
%-------------
%-------------
%-------------
%-------------

For a systematic  evaluation  one has  to insert  the ansatz  for 
$\gamma_{l,l+1}(t)$  into these expressions and has to separate different 
harmonics in the Josephson frequency $\omega$. A considerable simplification 
is  obtained  if one keeps  the  non-linear  effects  only  in the 
sin-term  of the Josephson  current  and linearizes the other terms with
respect to $\gamma$.
Then one obtains for the current density across the barrier with thickness b
\begin{equation}
j_{l,l+1} =j_c\sin\gamma_{l,l+1} + \sigma_0 {\hbar\over 2eb}\dot \gamma_{l,l+1} + 
\sigma_1 (\mu_{l} - \mu_{l+1})/b 
\end{equation}
and for the density response
\begin{equation}
\delta \rho_l = \chi^{(0)}_{\rho\rho} \mu_l + 
\chi^{(2)}_{\rho\rho}(\mu_{l+1}+\mu_{l-1}- 2\mu_{l}) 
+ {\hbar\over 2eb} \sigma_1 (\gamma_{l-1,l} - \gamma_{l,l+1})
\; .\end{equation}
In the normal state 
$\sigma_0=\sigma_1$ and $j_{l,l+1}$ depends only on the voltage across the
barrier. In the superconducting state $\gamma_{l,l+1}$ and $\mu_l$ have
separate physical meaning and we need the second equation for $\delta \rho_l$ to 
determine $\mu_l$ as a function of the voltage. The density reponse function 
$\chi^{(0)}_{\rho\rho}$ is finite in the superconducting state. At low 
temperatures it is only weakly frequency dependent. Approximately it is given 
by $\chi^{(0)}_{\rho\rho}\simeq -2e^2N_2(0)$,
where $N_2(0)$  is the two-dimensional density of states of the 
electron   gas   in   the   CuO$_2$-layers.   The   conductivities 
$\sigma_{0,1}$  as  well  as $\chi^{(2)}_{\rho\rho}$  describe  the 
charge exchange with the neighboring  layers and are proportional 
to $t^2_\perp$. 
Adding finally the displacement current we obtain a relation
between the electronic current density across the barrier with the 
external current density,
\begin{equation}
j= j_{\l,l+1} +  \epsilon\epsilon_0 \dot E_{l,l+1}
\; .\end{equation}

\begin{figure}
   \begin{center}
   \begin{tabular}{c}
   \psfig{figure=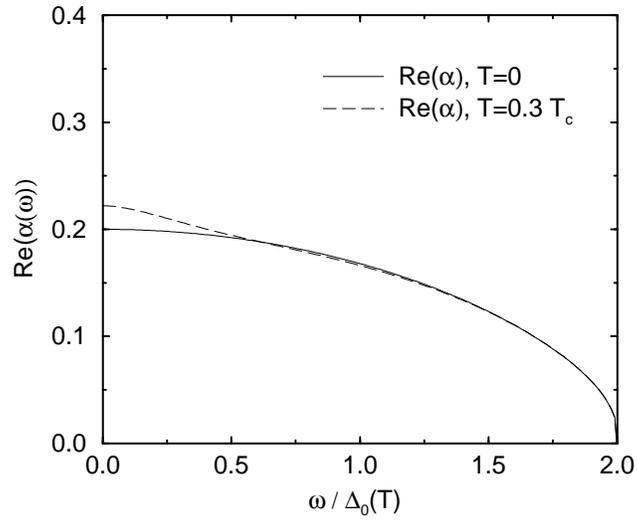,height=7.5cm} 
   \end{tabular}
   \end{center}
   \caption[example] 
   { \label{fig3}          
Real part of $\alpha$ as function of the frequency 
$\omega$ for a d-wave 
superconductor at two different temperatures $T$.
} 
\end{figure}

In a first  step  we eliminate  the scalar  potential  difference 
$\mu_{l}$  in  favour  of  the  gauge-invariant  phase  difference 
$\gamma_{l,l+1}$  and the electric  field  by using the Josephson 
relation 
\begin{equation}
{\hbar  \over 2e} \dot \gamma_{l,l+1}(t)  = b E_{l,l+1}  + 
\mu_{l+1} - \mu_l 
\; . \end{equation} 
In the next step we express the charge fluctuations with help  of the 
Maxwell  equation 
\begin{equation}
\delta \rho_l = \epsilon_0 \epsilon (E_{l,l+1} - E_{l-1,l}) 
\; . \end{equation}
We then finally  arrive  at the  following  differential 
equation for the phase:
\begin{equation}
{j\over j_c} = \left(1- \alpha \Delta^{(2)} \right) \sin \gamma_{l,l+1} + 
{1\over  \omega_c} \left(1- \eta \Delta^{(2)}\right) \dot \gamma_{l,l+1}
+ {1\over  \omega_p^2} 
\left( 1-\zeta \Delta^{(2)}\right) \ddot \gamma_{l,l+1}
\label{diffeq_gamma}
\end{equation}
where $\omega_p^2 = j_c2eb/(\epsilon\epsilon_0\hbar)$, 
$1/\omega_c=\sigma_0 /(\epsilon\epsilon_0\omega_p^2)$ 
The dimensionless  quantities  $\alpha, \eta, \zeta$ describe the 
coupling  of the phase-difference  in different  layers  via  the 
derivative operator $\Delta^{(2)}$, which is defined as 
$\Delta^{(2)} f_l= f_{l+1}+f_{l-1} - 2f_l$. In particular 
\begin{equation}
\alpha = -\epsilon\epsilon_0/(b \chi^0_{\rho\rho}) + O(t_\perp^2)
\; , \end{equation}
\begin{equation}
\eta = -\epsilon\epsilon_0/(b 
\chi^0_{\rho\rho}) (1- 2\sigma_1/\sigma_0) + O(t_\perp^2)
\; . \end{equation}
$\zeta$ is proportional to $t^2_\perp$.
For $\omega\ll\Delta$ the quantity $\alpha$ is only weakly 
frequency dependent 
(see Fig. \ref{fig3}). Therefore it will be approximated by 
its value at $\omega=0$, $\alpha(0)=\epsilon\epsilon_0/(2e^2 b N_2(0))$.  
If we neglect $\eta$ and $\zeta$ we arrive  back at the 
theory of Koyama and Tachiki \cite{Koyama96}. 
In fact at $\omega\ll T \ll\Delta$ we find  
$\sigma_1 \simeq \sigma_0/2$ for d-wave superconductors. 
Thus neglecting $\eta$ and $\zeta$ seems to be
a good approximation for small values of $\omega$. 
For strong coupling in $c$-direction one ends up with a situation, where charge
fluctuations on the layers are suppressed, $\delta\rho_l$  = 0, but a finite 
$\mu_l$ is generated by charge-imbalance of quasi-particles \cite{Ryndyk98}.
A study of these effects in this model will be done in the future. 
%-------------

\section{Influence on the current-voltage characteristics}
\label{sect:iv-charac}

\begin{figure}
   \begin{center}
   \begin{tabular}{c}
   \psfig{figure=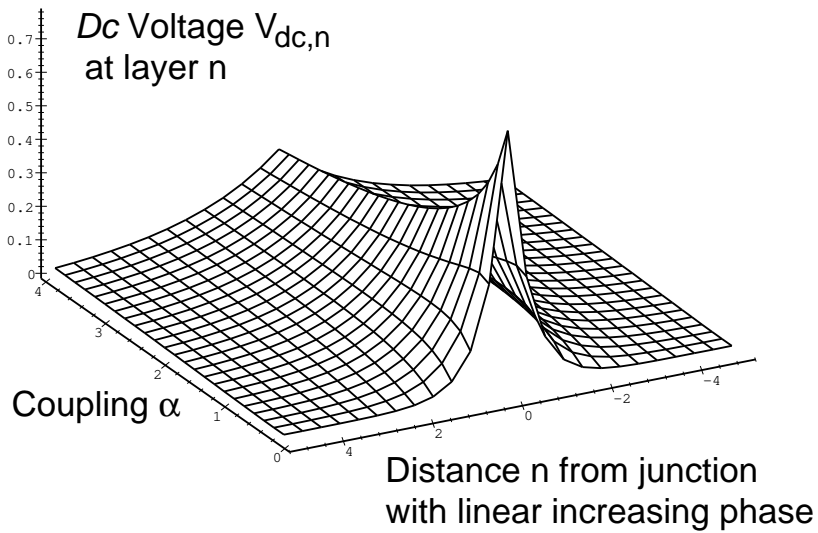,height=7cm} 
   \end{tabular}
   \end{center}
   \caption[example] 
   { \label{fig4}          
Distribution of the dc-voltage in the neighborhood of one resistive junction
as function of the distance $n$ and the coupling constant $\alpha$. 
} 
\end{figure}

In our further  discussion  of the experimental  implications  of 
this effect on the current-voltage characteristics we restrict  
ourselves  to the approximation (\ref{diffeq_gamma}) 
with $\alpha = const, \eta= \zeta=0$ 
supplemented by 
\begin{equation}
{\hbar\over 2e} \dot \gamma_{l,l+1} = V_{l,l+1} - \alpha 
(V_{l+1,l+2}+V_{l-1,l} - 2 V_{l,l+1})
\end{equation}
where we have defined the voltage $V_{l,l+1} = 
\int_l^{l+1} E_z dz = bE_{l,l+1}$.
In particular, we obtain a relation for the dc-voltages, if we replace $\dot
\gamma_{\l,l+1}$ by its time-average $<\dot \gamma_{l,l+1}>$. A junction is called
to be in the resistive state if $<\dot \gamma_{l,l+1}(t)> \ne 0$. 
In the case of one junction in the resistive state there is 
a finite voltage-drop also in the neighboring junctions due to the coupling
$\alpha$. This is shown in Fig. \ref{fig4}. Note that the total  
voltage  is given  by 
\begin{equation}
V= \sum_l V_{l,l+1} = {\hbar \over 2e} \sum_l \dot \gamma_{l,l+1}
\end{equation}
as in the absence of the coupling. 

%-------------
%-------------

We want  to study  in particular  the influence  of the dynamical  coupling 
between different barriers in the resistive state. This 
will  be  done  both  by  numerical  simulations  and  analytical 
calculations using a Green's function technique similar to that used by Takeno 
in a different context\cite{Takeno97}. Details of the calculations will be published elsewhere.

\begin{figure}
   \begin{center}
   \begin{tabular}{c}
   \psfig{figure=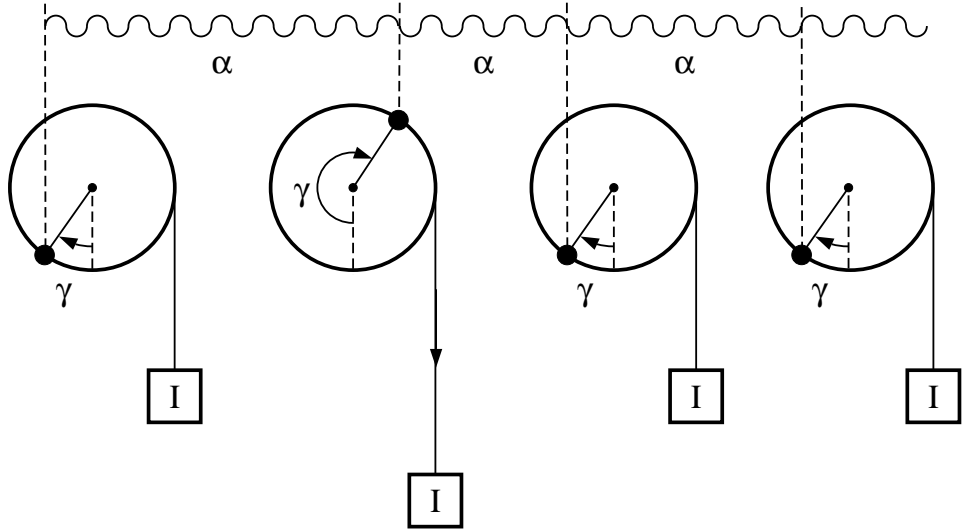,height=7cm} 
   \end{tabular}
   \end{center}
   \caption[example] 
   { \label{fig5}          
Mechanical analog for a stack of coupled Josephson junctions.
The angle $\gamma$ of a rotator corresponds to the phase difference 
of a josephson junction.
One phase is running (= resistive state), the other phases are only 
oscillating (= superconducting state). 
} 
\end{figure}

\begin{figure}
   \begin{center}
   \begin{tabular}{c}
   \psfig{figure=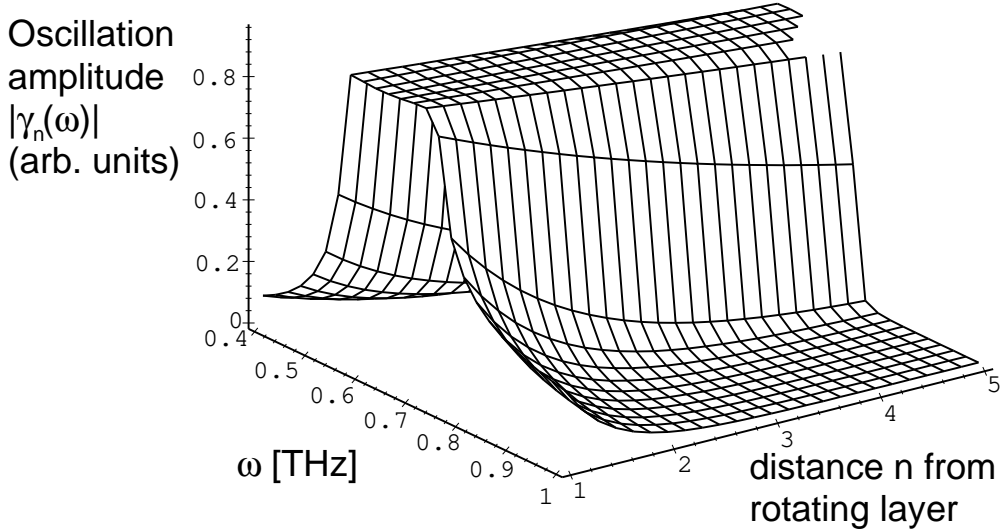,height=7cm} 
   \end{tabular}
   \end{center}
   \caption[example] 
   { \label{fig6}          
Oscillation amplitude of the phase $|\gamma_n(\omega)|$ 
in the different superconducting barriers 
as function of the distance $n$ from the resistive junction and as function
of the oscillation frequency $\omega$. 
} 
\end{figure}

Let us begin with the discussion  of one barrier in the resistive 
state.  It may be helpful  to visualize  the dynamics  of such a 
system by considering the mechanical analog of the RSJ model: The 
dynamics of each phase difference can be described by a pendulum 
with  a constant  torque  being proportional  to  the  bias 
current $I$ which is the same in all barriers (Fig. \ref{fig5}). 
In the absence of 
the coupling $\alpha$  the pendulum is either rotating,  this 
corresponds  to the  resistive  state  of the  barrier,  or has a 
constant  phase, this corresponds  to the superconducting  state. 
In the presence  of the coupling  which produces 
an  additional  torque, one still  can distinguish 
between  these  two  types  of motion:   a rotating  state  and a 
non-rotating  vibrating  state.  In the rotating  state there is a 
running  phase, $\gamma(t) = \gamma_0 + \omega t + \delta\gamma(t)$
with a finite  value  of the Josephson frequency $\omega = <\dot 
\gamma>$.   In the 
non-rotating  state  $<\dot  \gamma>=  0$, but there are still
oscillations. Such localized solutions are known in non-linear dynamics 
as roto-breathers \cite{Flach98}. The running  phase  in the  resistive 
barrier  causes  finite  phase  oscillations  in the  neighboring 
barriers.   The amplitude  of these  oscillation  
depends  on the 
ratio  of  the  rotation  frequency, i.e. the  Josephson  frequency 
$\omega$, and the eigenfrequencies of the oscillations of the coupled 
system  at small amplitudes.  The latter are determined by the 
Josephson   plasma  frequency $\omega_p$ and have a  bandwidth   which  is 
proportional   to  the  coupling  $\alpha$.    
In  Fig. \ref{fig6} we  show  the  result  of an analytical  
calculation of the oscillation amplitude for different barriers 
as a function of the 
distance from the resistive barrier and as a function of the 
Josephson frequency  $\omega$.  
In this example the plasma frequency is $\omega_p/(2\pi) = 0.6$ THz and  
$\alpha=0.2$.
If  the  rotation 
frequency  is  high, $\omega \gg \omega_p$, which is usually the case  
for  intrinsic 
Josephson systems, the oscillation  amplitude in the neighboring 
barriers   is  small  and  falls   off  exponentially   with  the 
distance  from the resistive  barrier.   In the case of a 
Josephson  frequency $\omega$ within the plasma band, the phase-rotation 
in one barrier leads to long-range plasma oscillations 
in    the    neighborhood.    These    are    the    longitudinal 
plasma-oscillations   considered  by  several  authors  which  in 
principle can also be excited by longitudinal  electric fields in 
the purely superconducting state. 

%-------------

%-------------
%-------------

\begin{figure}
   \begin{center}
   \begin{tabular}{c}
   \psfig{figure=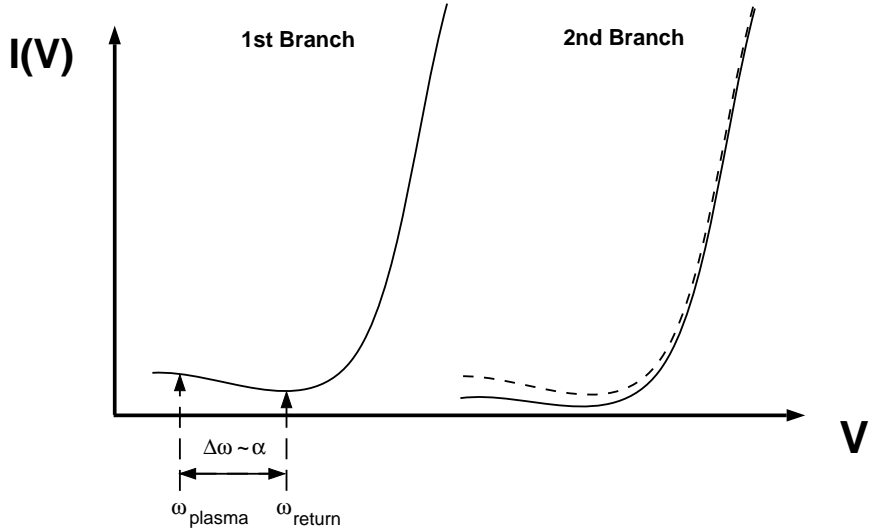,height=7cm} 
   \end{tabular}
   \end{center}
   \caption[example] 
   { \label{fig7}          
Schematic plot of the current-voltage characteristics for two coupled 
junctions (solid line) in comparison with two uncoupled junctions
(dashed line)
} 
\end{figure}

From an experimental  point  of view it is important  to know how 
this effect influences the current-voltage  characteristics of an 
intrinsic  Josephson  system, 
showing a multi-branch structure, where the n-th branch corresponds to 
n resitive barriers. 
As we have seen above,  the total 
voltage measured  for the stack of Josephson  junctions  is still 
given  by the sum of the Josephson  frequencies  of the resistive 
barriers. For the total dc-voltage only the values of 
$<\dot \gamma_{l,l+1}>$ for the resistive barriers contribute.  
In  the  case  of  two  or more  resistive  barriers in the stack 
it makes  a 
difference  whether these resistive barriers are next neighbors or 
are separated by one or more non-resistive junctions.  
We have checked  this both  by numerical 
simulations  and analytical  calculations.   We find  that in the 
case of two resistive  barriers  next to each other  the rotation 
frequency  of both  junctions  is the same,   but (at the same current)
is slightly  higher than in the case of well-separated  junctions.  
For two uncoupled junctions, of course,  
the total voltage is just the double of one 
resistive junction (the first branch).
This is shown in Fig. \ref{fig7} 
where  we compare  the second  branch  of the  current  voltage 
characteristics for two neighboring resistive  junctions 
with the total voltage of two well separated resistive junctions. 
For realistic values of $\alpha < 1$ 
the effect  is very small,  therefore  we have exaggerated  in the 
figure this effect.  More pronounced  is the difference  near the 
plasma  frequency: If we lower  the  voltage 
towards  the upper edge of the plasma band, at least one of the resistive  
junctions  returns to 
the superconducting  state.  This return  point (the minimum of the curves)  
is different  for 
the two situations: for coupled junctions the return-voltage and the
dc-current is smaller than for uncoupled resistive junctions. 

%-------------
%--------------

Our numerical  simulations  also  show  that  in the case  of two 
neighboring resistive junctions the two phases rotate coherently. 
This  remains  true  even  if the  critical  currents  of the  two 
junctions  are slightly  different.   This phase-locking  is very 
important for application  of coupled Josephson oscillations  for 
the generation or amplification of radiation.

Our experiments  on the intrinsic  Josephson  systems BSCCO and TBCCO 
show a very precise additive structure of the different  branches 
at least for the lower branches.   Deviations  from the additivity 
at higher order branches  may be attributed  to heating  effects. 
This poses an upper  limit to the coupling  parameter  of $\alpha  
< 1$. An upper limit of $\alpha$ is also obtained from the hysteretic return
point of the current voltage characteristics from the first resistive branch
to the superconducting state. A lower limit of this return point 
is given by the upper edge of the plasma band given by
$\omega_p\sqrt{1+2\alpha}$. From the experimental value for $\omega_p$ and 
the return voltage $\omega_{return}$ one can also estimate $\alpha<1$. 
Finally one may calculate $\alpha$ for a
2-dimensional electron gas: for $\epsilon\approx 25$ one obtains $\alpha= 0.1$.

In this paper we have studied the coupling of Josephson oscillations in
different barriers due to charge fluctuations. Another mechanism which also
leads to such a coupling is the excitation of $c$-axis phonons in the stack by 
Josephson oscillations, which we have studied recently \cite{Helm97}. 
The phonons
excitated in one resistive barrier induce field oscillations also in
neighboring barriers and thus support coherence of Josephson oscillations.

\section{Summary}
\label{sect:summary}

Starting from a microscopic  model for the Josephson  effect in a 
stack  of  superconducting  layers  coupled  by tunneling 
barriers  we  have  derived  a set  of  equations  for  the 
gauge-invariant   phase  difference in different  barriers. 
A simplified model has then been used to 
study  the  non-linear   dynamics  of  the  coupled  system.   In 
particular,  we have  studied  the difference  in the  current-voltage 
characteristics   of  two  neighboring   resistive   junctions  in 
comparison  with well separated resistive junctions.   From the regular 
additive  structure  of the current-voltage  branches  found in 
experiments  on BSCCO and TBCCO and the return voltage one can derive an upper limit of 
the coupling constant of the order of $\alpha < 1$.

\acknowledgments
This work was supported by the Bayerische Forschungsstiftung (Ch.P.), the
German Science Foundation (Ch.H.), and the Humboldt-Foundation (A.S.).

   \bibliography{paper2}
   \bibliographystyle{spiebib}   %>>>> makes bibtex use spiebib.bst

\end{document}